\documentstyle[amsmath,amssymb,graphicx]{article}

\def\be{\begin{eqnarray}}
\def\ee{\end{eqnarray}}
\def\nn{\nonumber}

\def\p{\partial}

\hoffset= - 1.0in         

\title{{\bf Nekrasov Functions from Exact BS Periods:\\
the Case of $SU(N)$
} \vspace{.2cm}}
\author{{\bf A.Mironov}\footnote{ {\small {\it
Lebedev Physics Institute} and {\it ITEP, Moscow, Russia}};
mironov@itep.ru; mironov@lpi.ru} \ and {\bf
A.Morozov}\thanks{{\small {\it ITEP, Moscow, Russia}};
morozov@itep.ru} \date{ }}

\begin{document}

\maketitle

\vspace{-6.0cm}

\begin{center}
\hfill FIAN/TD-27/09\\
\hfill ITEP/TH-58/09\\
\end{center}

\vspace{4cm}

\begin{abstract}
In \cite{NSBS2} we suggested that the Nekrasov
function with one non-vanishing deformation parameter
$\epsilon$ is obtained by the standard Seiberg-Witten
contour-integral construction.
The only difference is that the Seiberg-Witten differential
$pdx$ is substituted by its quantized version
for the corresponding integrable system, and contour
integrals become exact monodromies of the wave function.
This provides an explicit formulation of the earlier
guess in \cite{NeSha}.
In this paper we successfully check this suggestion
in the first order in $\epsilon^2$ and
the first order in instanton expansion for the $SU(N)$ model,
where non-trivial is already consistency
of the so deformed Seiberg-Witten equations.
\end{abstract}

\section{Introduction}

Integrability plays a very important role in modern
theoretical physics, because effective actions of
quantum theories always exhibit integrability properties
\cite{UFN2}.
The basic reason for this is the freedom to change
integration variables in functional integral.
If this freedom is preserved on some "mini-superspace"
(moduli space) of coupling constants, the universality
classes of effective actions are labeled by some simple
and well known integrable system in low space-time dimensions.
Today there is a number of interesting examples where
this phenomenon manifests itself.
One of them is the Seiberg-Witten (SW) theory,
describing the low-energy effective actions of $4d$ ${\cal N}=2$
supersymmetric gauge theories \cite{SW}: universality
classes in this case are labeled by $1d$
integrable systems \cite{GKMMM}
like Toda \cite{GKMMM,Toda}, Calogero \cite{Cal},
Ruijsenaars \cite{Ruij}
models and spin chains \cite{spin}.
An alternative description of the SW theory is in terms of the Nekrasov
functions \cite{Nek}, which originally appeared from an
attempts to perform a regularized integration \cite{LNS}
over instanton moduli spaces with the help of
Duistermaat-Heckman (localization) technique \cite{DH}.
Today the Nekrasov functions have become an important class of special
functions in string theory \cite{sfst}, generalizing the ordinary
hypergeometric series in a non-trivial way \cite{MMnf},
and the AGT conjecture \cite{AGT} implies that they provide a good
starting point to describe at least the entire set of $2d$
conformal blocks.
All this makes description of the Nekrasov functions in terms of
integrability theory an important and urgent problem.
Of course, from the general perspective, the Nekrasov functions
are fragments of KP-Toda $\tau$-functions, closely related
to discrete matrix models \cite{dimamo} and combinatorics of
symmetric groups \cite{sygr}.
However, their relevance for the SW theory implies that there should
be  relation to a much simpler class of $1d$ integrable systems.
A first guess in this direction was made in a recent paper \cite{NeSha},
where it was suggested that introducing the $\epsilon$ parameters
corresponds in some way to a direct quantization of the integrability/SW
relation of \cite{GKMMM}\footnote{A similar relation of $2d$ supersymmetric theories
and quantum integrable systems can be found in \cite{NeShap}.}. 
In \cite{NSBS2} we provided an explicit
description of this quantization procedure.

The SW theory \cite{SW} defines a {\it prepotential} ${F}_{SW}(\vec a)$
from the system of equations
\be
a_i = \oint_{A^i} dS^{(0)} = \Pi^{(0)}_{A_i}, \nn \\
\frac{\p {F_{SW}(\vec a)}}{\p a_i} = \oint_{B_i}  dS^{(0)} = \Pi^{(0)}_{B_i}
\label{SWeq}
\ee
where contour integrals are the Bohr-Sommerfeld (BS) periods of an associated
$1d$ integrable system \cite{GKMMM}.
The claim of \cite{NSBS2} is that Nekrasov's prepotential
${\cal F}(\vec a|\epsilon_1)$ with one $\epsilon$-parameter
switched on
(in principle there can be arbitrary many such $\epsilon$-parameters,
though \cite{Nek} discusses just two)
is defined by {\it the same} system (\ref{SWeq}), only the BS
presymplectic differential $dS^{(0)} \approx \vec p d\vec q$
is substituted by its exact quantum counterpart: the one which
defines the phase of exact wave function of the integrable system.
To emphasize that the relevant moduli $\vec a$ are now different (deformed),
we rewrite this system in the slightly different notation:
\be
\alpha_i = \oint_{A^i} dS = \Pi_{A_i}, \nn \\
\frac{\p {\cal F(\vec\alpha|\epsilon)}}{\p \alpha_i} = \oint_{B_i}  dS
= \Pi_{B_i}
\label{qSWeq}
\ee
The deformed BS periods are no nothing but (Abelian) monodromies of the wave
function.

In \cite{NSBS2} we explicitly checked this suggestion (in the lowest orders
of various expansions) only in
the simplest $SU(2)$ case, when the relevant integrable system is
the ordinary sine-Gordon.
Though generalizations to $SU(N)$ Toda systems are well known to be
straightforward, this is an important check to be done, because for $N>2$
the system  (\ref{qSWeq}) could be non-resolvable: not any set of periods
can be represented as a gradient of something.
Consistency of the system can not be proved with the help of ordinary
Riemann's theorem $T_{ij} = T_{ji}$ as in the case of original SW theory,
already because, after the deformation,
$dS$ is no longer a SW differential with the property
$\delta(dS) = holomorphic$.
Still, a memory of the spectral Riemann surface survives
(it actually gets modified only in the vicinity of ramification points),
and we gave a technical argument at the
end of \cite{NSBS2} in favor of the consistency of (\ref{qSWeq}), and now
we are going to check that this system is indeed consistent and, moreover,
has ${\cal F}(a|\epsilon_1)$ as its solution. Like \cite{NSBS2}, we
are going to make this check only in the first orders of expansions in
$\epsilon_1^2$ and $\Lambda^{2N}$, and even this calculation is rather
cumbersome. A better proof should, of course, be searched for.

To simplify the calculations we exploit the existing knowledge about the SW
theory and the Nekrasov functions  as much as possible.
Actually we proceed in the following three steps.

\paragraph{Step 1. SW periods $\Pi^{(0)}$ and Nekrasov functions.}
The $SU(N)$ universality class of SW theory is labeled by a polynomial
\be
K(p) = \sum_{k=0}^N u_k p^k = u_N\prod_{i=1}^N(p-\lambda_i)
\ee
The SW/Toda spectral curve is given then by
\be
K(p) + \gamma\cos\phi = 0, \ \ \ \ \gamma = \Lambda^N,
\label{quacleq}
\ee
and the SW differential is
\be
dS^{(0)} = p\,d\phi
\ee
The periods $\Pi^{(0)}$ can be calculated in various ways, either directly
or with the help of the Picard-Fucks equations. We, however, take the
most economic and transparent way: we calculate $a_i(\vec\lambda)$
directly from the definition, but take the difficult dual periods
from the Nekrasov function
\be
F(\vec a) = \lim_{\epsilon_1,\epsilon_2\rightarrow 0}\epsilon_1\epsilon_2
\log Z_{LNS}(\vec a|\epsilon_1,\epsilon_2)
\ee

\paragraph{Step 2. WKB theory and deformed differential $dS$.}
The deformed differential $dS$ is an exact solution to the deformed (quantized)
equation (\ref{quacleq}), see the very last formula in \cite{NSBS2}:
\be
\left\{K\left(-i\hbar\frac{\p}{\p\phi}\right) + \gamma\cos\phi\right\}
\exp\left( \frac{i}{\hbar}\int^\phi \!\!dS\right) = 0
\label{Shreq}
\ee
and actually $\hbar = \epsilon_1$.
WKB theory \cite{WKB} provides an expansion of $dS = \sum_{k=0}^\infty
\hbar^k P_kd\phi$, where $P_0=p(\phi)$ is the "classical" momentum, that is, 
the root of (\ref{quacleq}), which is single-valued on the
spectral Riemann surface.
A technically reasonable way to calculate the periods of $P_kd\phi$
with $k>0$ is to represent $dS = \hat{\cal O}\, dS^{(0)}$ as an action of
some differential operator ${\cal O}$ (acting on parameters $u_i$ and $\gamma$):
then $\Pi_C = \hat{\cal O}\,\Pi^{(0)}_C$.

\paragraph{Step 3. The check of the "exact BS" suggestion of \cite{NSBS2}.}
Finally one
\begin{itemize}
\item Evaluates the deformed $A$-periods $\vec\alpha(\vec\lambda) =
\hat{\cal O}\!\left[\vec a(\vec\lambda)\right]$,
\item Substitutes these deformed $A$-periods into
the $\alpha$-derivatives of the known Nekrasov function
${\cal F}(\vec\alpha|\epsilon_1) = \lim_{\epsilon_2\rightarrow 0}
\epsilon_1\epsilon_2 \log Z_{LNS}(\vec \alpha|\epsilon_1,\epsilon_2)$,
\item Compares the result with deformed $B$-periods, obtained at step 1
from the $a$-derivatives of the SW-Nekrasov function
$F_{SW}(\vec a) = {\cal F}(\vec a|\epsilon_1=0)$, i.e. with
$\hat{\cal O}\left[\p_{\vec a}F_{SW}(\vec a)\right]$.
\end{itemize}

In other words, we are going to prove the relation
\be
\boxed{
\Pi_B \Big(\hat{\cal O} \left[\Pi^{(0)}_A(\lambda)\right] \Big)
= \hat{\cal O} \left[\Pi^{(0)}_B\Big(\Pi^{(0)}_A(\lambda)\Big)\right]
}
\label{comp}
\ee
extracting $\Pi_B^{(0)}(a)$ and $\Pi_B(a)$ from the Nekrasov functions
with vanishing and non-vanishing $\epsilon_1$ respectively,
explicitly evaluating $\Pi^{(0)}_A(\lambda)$ and deriving
operator $\hat{\cal O}$ from WKB theory.

All these steps are actually easily computerized and higher-order corrections
can be also analyzed after that. In this letter, however, we present as
many as possible formulas explicitly, without appeal to computer calculations.
In fact, there is a close similarity between emerging formulas and those
familiar from various matrix-model calculations, especially from \cite{AMM}
and the theory of CIV-DV potentials \cite{CIVDV}.

We actually begin in s.\ref{WKB} from step 2,
then proceed to step 1 in ss.\ref{NF} and \ref{per} and end
with step 3 in s.\ref{compa}.

\section{WKB theory and deformed differential $dS$
\label{WKB}}

\subsection{Conjugation of the differential operator }

\be
e^{-\frac{i}{\hbar}\int^x Pdx}
(-i\hbar\p)^n e^{\frac{i}{\hbar}\int^x Pdx}
= P^n - i\hbar\frac{n(n-1)}{2}P^{n-2}\dot P - \nn \\
- \hbar^2\left(
\frac{n(n-1)(n-2)}{6}P^{n-3}\ddot P + \frac{n(n-1)(n-2)(n-3)}{8}
P^{n-4}\dot P^2\right) + O(\hbar^3)
\label{difexp}
\ee
where $\dot P \equiv \p P$, while prime is reserved for
$P$-derivatives of $P$-dependent functions, see below.

\subsection{Shr\"odinger equation (\ref{Shreq}) for the
differential $dS$}

For
\be
K(z) = \sum_{k=0}^N u_k z^k
\label{Kdef}
\ee
one needs to solve
\be
\Big(K(-i\hbar\p) + \gamma\cos x \Big)
e^{\frac{i}{\hbar}\int^x Pdx} = 0
\ee
Making use of (\ref{difexp}), this can be rewritten as
\be
K(P) - \frac{i\hbar}{2}K''(P)\dot P - \hbar^2\left(
\frac{1}{6}K'''(P)\ddot P + \frac{1}{8}K''''(P)\dot P^2
\right) = -V(x) = -\gamma\cos x
\ee
Substituting
\be
P = p + \hbar P_1 + \hbar^2 P_2 + O(\hbar^3),
\ee
one gets
\be
K(p) = -V(x), \nn \\
P_1 = -i\frac{K''(p)\dot p}{2K'(p)}
= -\frac{i}{2}\p \Big(\log K'(p)\Big), \nn \\
P_2 =
\left(\frac{3K''^3}{8K'^3}-\frac{K''K'''}{2K'^2}
+ \frac{K''''}{8K'}\right)\dot p^2
+ \left(-\frac{K''}{4K'^2} + \frac{K'''}{6K'}\right)\ddot p, \nn \\
\ldots
\ee
Here and below $K$ with omitted argument denotes $K(p)$,
similarly $K' = K'(p)$ and so on.

From the first equation it follows that
\be
\dot p = -\frac{V'}{K'}, \nn \\
\ddot p = -\frac{V''}{K'} - \frac{K''V'^2}{K'^3}, \nn \\
\ldots
\ee
and
\be
P_2 = \left(\frac{K''}{4K'^3} - \frac{K'''}{6K'^2}\right)V''
+ \left(\frac{5K''^3}{8K'^5} -\frac{2K''K'''}{3K'^4}
+ \frac{K''''}{8K'^3}\right)V'^2
\label{P2}
\ee

\subsection{Simplified expression for contour integrals}

For contour integrals integration by parts is allowed,
and this allows one to considerably simplify the integral of (\ref{P2}):
\be
\Pi^{(2)}_C \equiv
\hbar^2\oint_C P_2dx = \frac{\hbar^2}{24}\oint_C \left(
\frac{K''^2}{K'^3}-\frac{K'''}{K'^2}\right)V'' dx
\ee
For $K(p) = \frac{1}{2}p^2-E$, these formulas reproduce
the standard WKB expressions used in \cite{NSBS2}.

\subsection{Exact periods from BS periods and the
operator $\hat{\cal O}$}

For $V(x) = \gamma\cos x$ one has $V''=-V$.
Further, from $K(p) = -V = -\gamma\cos x$ and (\ref{Kdef})
it follows that
\be
\gamma\frac{\p p}{\p\gamma} = -\frac{V}{K'},\nn\\
\frac{\p p}{\p u_j} = - \frac{p^j}{K'}, \nn \\
\gamma\frac{\p^2 p}{\p\gamma\p u_j} = -
\left(\frac{K''}{K'^3}p^j - \frac{jp^{j-1}}{K'^2}\right)V
\ee
and
\be
\frac{\hbar^2\gamma}{24}\frac{\p}{\p\gamma}\left(
\sum_j j(j-1)u_j\frac{\p}{\p u_{j-2}}\right) p
= -\frac{\hbar^2}{24}\left(
\frac{K''^2}{K'^3}-\frac{K'''}{K'^2}\right)V
\ee
This means that for any closed contour $C$
\be
\boxed{
\Pi^{(0)}_C + \Pi^{(2)}_C = \hat {\cal O}\,\Pi^{(0)}_C =
\left(1+\frac{\hbar^2\gamma}{24}\frac{\p}{\p\gamma}
\sum_j j(j-1)u_j\frac{\p}{\p u_{j-2}}\right)\Pi^{(0)}_C
}
\label{OPO}
\ee

\section{Nekrasov functions
\label{NF}}

The Nekrasov functions are now reviewed in numerous papers \cite{nfrev}.
They are obtained from the LNS contour multi-integrals \cite{LNS},
which in the simplest $SU(N)$ case look like
\be
Z_{LNS}(\vec a|\{\epsilon\}) \equiv \sum_k {1\over k!}\left(
{\epsilon\over\epsilon_+\epsilon_-}\right)^k
\prod_{I=1}^k \oint {d\varphi_I\over 2\pi i}
\frac{Q(\varphi_I)}{\prod_{j=1}^N(\varphi_I-a_j)(\varphi_I-a_j+\epsilon)}
\prod_{I<J}^N \frac{ \varphi_{IJ}^2
\prod_{a<b}\Big(\varphi_{IJ}^2-(\epsilon_{a}+\epsilon_b)^2\Big)\ldots}
{\prod_a \Big(\varphi_{IJ}^2-\epsilon_a^2\Big)
\ldots}
\ee
where the polynomial $Q$ depends on the matter content of the model,
for pure gauge theory $Q(\varphi)=\Lambda^{2N}$.
The crucial step was done in \cite{Nek}:
the integral was rewritten as an explicit sum over a
collection of Young diagrams, which provided a practically
useful expansion basis for various purposes.

The Nekrasov function for $SU(N)$ is given by
\be
{\cal F}(a|\epsilon_1) = {\cal F}^{pert}(a|\epsilon_1)
+ {\cal F}^{inst}(a|\epsilon_1)
\label{NFfull}
\ee
where the perturbative contribution for $\epsilon\neq 0$
looks nice only when the $a$-derivative is taken,
\be
-\frac{\p {\cal F}^{pert}}{\p a_i} = 2\epsilon_1
\sum_{j\neq i} \log \frac{\Gamma(1+a_{ij}/\epsilon_1)}
{\Gamma(1-a_{ij}/\epsilon_1)} =
\sum_{j\neq i} 4a_{ij}\left\{
\left(\log\frac{a_{ij}}{\Lambda}-1\right)
+ \sum_{m=1}^\infty \frac{B_{2m}}{2m(2m-1)}
\left(\frac{\epsilon_1}{a_{ij}}\right)^{2m}\right\}
= \nn \\ =
4\sum_{j\neq i} \left\{ a_{ij}
\left(\log\frac{a_{ij}}{\Lambda}-1\right)
+\frac{\epsilon_1^2}{12a_{ij}}
+ O(\epsilon_1^4)\right\}
\label{NFpexpan}
\ee
while the instanton part is a series in powers of $\gamma^2=\Lambda^{2N}$,
of which we will need only the first term
(associated with the single-box Young diagrams)
\be
{\cal F}^{inst} = \frac{\Lambda^{2N}}{2u_N^2}
\sum_{i=1}^N \prod_{j\neq i}\frac{1}
{a_{ij}(a_{ij}+\epsilon)}
+ O(\Lambda^{4N}) = \nn \\ = \frac{1}{2u_N^2}
\sum_{i=1}^N \frac{\Lambda^{2N}}
{\prod_{j\neq i} a_{ij}^2}\left\{1 + \epsilon^2
\left(\sum_{j\neq i}\frac{1}{a_{ij}^2} + \sum_{j<k}
\frac{1}{a_{ij}a_{ik}}\right) + O(\epsilon^4)\right\}
+ O(\Lambda^{4N})
\label{NFiexpan}
\ee

The SW prepotential $F_{SW}(\vec a)$ is defined by the same
formulas, only all terms with $\epsilon^2$ are omitted,
see s.\ref{BSBnf} below.

\section{SW/BS periods $\Pi^{(0)}$
\label{per}}

As explained in the Introduction, we evaluate the $A$ periods
$a_i = \Pi^{(0)}(A^i)$ as functions of $\lambda_i$ and $\gamma$
directly, while
the $B$ periods $\Pi^{(0)}(B_i)$ will be obtained from
(\ref{SWeq}) by differentiating $F_{SW}(a_i)$ from the
previous section and then substituting there $a_i(\vec\lambda)$.

\subsection{SW/BS $A$-periods $\vec a$ through the roots $\vec\lambda$}

Shifting $\phi \rightarrow \phi - iN\log\Lambda$ in (\ref{quacleq}),
one obtains
\be
e^{i\phi} = -\Big(2K(p) + \Lambda^{2N}e^{-i\phi}\Big)
= -2K(p)\left(1 - \frac{\Lambda^{2N}}{4K(p)^2}\right)
\ee
Therefore,
\be
\Pi^{(0)} = i\oint p d\phi = \oint \frac{pdK}{K}
+ \frac{\Lambda^{2N}}{2}\oint \frac{pdK}{K^3}
= \sum_k \oint \frac{pdp}{p-\lambda_k}
+ \frac{\Lambda^{2N}}{4u_N^2}\oint \frac{dp}{\prod_k (p-\lambda_k)^2}
\ee
and
\be
\boxed{
a_i = \Pi^{(0)}_{A_i} = \lambda_i - \frac{\Lambda^{2N}}{2u_N^2
\prod_{k\neq i}\lambda_{ik}^2}\sum_{k\neq i}\frac{1}{\lambda_{ik}}
}
\label{avsla}
\ee

\subsection{SW/BS $B$-periods from Nekrasov function
\label{BSBnf}}

Putting $\epsilon=0$ in formulas from s.\ref{NF} one obtains
\be
\Pi^{(0)}_{B_i} =
-\frac{1}{4}\frac{\p{\cal F_{SW}}}{\p a_i} =
\sum_{j\neq i} a_{ij}\left(\log\frac{a_{ij}}{\Lambda}-1\right)
-\frac{\Lambda^{2N}}{8u_N^2}\frac{\p}{\p a_i}
\sum_{j=1}^N \frac{1}{\prod_{k\neq j} a_{jk}^2} + O(\Lambda^{4N}) = \nn \\
= \sum_{j\neq i} a_{ij}\left(\log\frac{a_{ij}}{\Lambda}-1\right)
+\frac{\Lambda^{2N}}{4u_N^2}\left(\frac{1}{\prod_{k\neq i} a_{ik}^2}\sum_{k\neq i}
\frac{1}{a_{ik}} + \sum_{j\neq i}\frac{1}{a_{ij}^3\prod_{k\neq i,j}a_{jk}^2}\right)
\label{0Bvsa}
\ee

\subsection{BS $B$-periods through the roots $\vec\lambda$}

In order to apply operator $\hat{\cal O}$, one needs the periods
expressed through the roots $\vec\lambda$ or coefficients $\vec u$
rather than through the moduli $\vec a$.
Thus, one needs to substitute $\vec a(\vec\lambda)$ from
(\ref{avsla}) into (\ref{0Bvsa})
\be
\boxed{
\Pi^{(0)}_{B_i} = \sum_{j\neq i} a_{ij}(\lambda)\left(\log\frac{\lambda_{ij}}{\Lambda}-1\right)
+\frac{\Lambda^{2N}}{4u_N^2}\left(\frac{1}{\prod_{k\neq i} \lambda_{ik}^2}\sum_{k\neq i}
\frac{1}{\lambda_{ik}} + \sum_{j\neq i}\frac{1}{\lambda_{ij}^3\prod_{k\neq i,j}\lambda_{jk}^2}\right)
}
\label{0Bvsla}
\ee
In the one-instanton approximation the only difference between
(\ref{0Bvsla}) and (\ref{0Bvsa}), except for a simple substitution
$a_i\rightarrow \lambda_i$, is that the coefficient in front
of logarithm is now $a_{ij}$, not $\lambda_{ij}$. The change of
logarithm's argument does not contribute.

\section{Quantized SW prepotential and Nekrasov function
\label{compa}}

We are now ready to act with operator (\ref{OPO}),
\be
\hat{\cal O}
=\left(1+\frac{\hbar^2\gamma}{24}\frac{\p}{\p\gamma}
\sum_j j(j-1)u_j\frac{\p}{\p u_{j-2}}+ O(\hbar^4)\right)
= 1 + \frac{\epsilon_1^2}{24}\hat{\cal O}^{(2)} + O(\epsilon^4)
\ee
on (\ref{avsla}) and (\ref{0Bvsla}), substitute the former one
into the full Nekrasov function (\ref{NFfull})-(\ref{NFiexpan})
and compare its derivative with the latter one.
The results coincide, thus validating the suggestion of
\cite{NSBS2} in the first order in $\Lambda^{2N}$ and $\epsilon^2$.

\subsection{Specifics of the second-order approximation}

Operator $\hat {\cal O}^{(2)}$ acts only on the $\Lambda$-dependent
($\gamma = \Lambda^N$) quantities, and the $u$-differential operator
can be conveniently expressed through the $\lambda$-derivatives:
\be
\sum_{j=0}^N j(j-1)u_j\frac{\p}{\p u_{j-2}} =
-\sum_{m=1}^N \frac{K''(\lambda_m)}{K'(\lambda_m)}\frac{\p}{\p\lambda_m}
\ee
It can be easily tested by acting on $p(u_i)$ and using $K'(p)\frac{\p p}{\p u_j} = -p^j$.

Identity (\ref{comp}), which we want to prove, in the leading
approximation can be rewritten as follows.
Its left hand side is
\be
\Pi_{B_i}\left(\vec a + \frac{\epsilon_1^2}{24}\hat{\cal O}^{(2)}[\vec a]\right) =
\Pi_{B_i}^{(0)}(\vec a)
+  \frac{\epsilon_1^2}{24}\sum_{j=1}^N
\hat{\cal O}^{(2)}[a_j(\vec\lambda)]\frac{\p}{\p a_j}\Pi^{(0)}_{B_i}(\vec a) + \nn \\
+ \frac{\epsilon^2}{24}\left(
2\sum_{j\neq i}\frac{1}{a_{ij}} +
\frac{12\Lambda^{2N}}{u_N^2\prod_{j\neq i}a_{ij}^2}\left(
\sum_{j\neq i}\frac{1}{a_{ij}^2} + \sum_{j<k}\frac{1}{a_{ij}a_{ik}}\right)
\right)
\ee
while its right hand side is
\be
\Pi_{B_i}^{(0)}(\vec a) + \frac{\epsilon_1^2}{24}\hat{\cal O}^{(2)}\left[
\Pi_{B_i}^{(0)}\big(\vec a(\vec\lambda)\big)\right]
\ee
Thus what we prove in this letter is
\be
\hat{\cal O}^{(2)}\left[\Pi_{B_i}^{(0)}\big(\vec a(\vec\lambda)\big)\right]
-\sum_{j=1}^N
\hat{\cal O}^{(2)}[a_j(\vec\lambda)]\frac{\p}{\p a_j}\Pi^{(0)}_{B_i}(\vec a)
= 2\sum_{j\neq i}\frac{1}{a_{ij}} +
\frac{12\Lambda^{2N}}{u_N^2\prod_{j\neq i}a_{ij}^2}\left(
\sum_{j\neq i}\frac{1}{a_{ij}^2} + \sum_{j<k}\frac{1}{a_{ij}a_{ik}}\right)
\label{comp2}
\ee
In the next subsection we explicitly describe the check for $\Lambda$-independent
terms in this formula.
The single-instanton contributions, i.e. the terms with $\Lambda^{2N}$,
also match at both sides, but formulas are somewhat lengthy and we do
not present them in this letter.

\subsection{Perturbative level}

For the perturbative part of the Nekrasov function
the difference between $\vec a$ and $\vec\lambda$ is inessential.
The $\hbar$-corrections ($\hbar=\epsilon_1$) to the $\Lambda$-independent
piece in ${\cal F}(\vec a|\epsilon_1)$ arise from
the action of deformation operator $\hat{\cal O}$ on
the logarithm in perturbative part of the SW prepotential,
\be
-\hat{\cal O}\ \frac{\p {\cal F}}{\p a_i} =
\left(1 + \frac{\hbar^2}{24}\gamma\frac{\p}{\p\gamma}
\sum_k k(k-1)u_k\frac{\p}{\p u_{k-2}} + \ldots\right)
\sum_{j\neq i} 4 a_{ij}\log\frac{a_{ij}}{\Lambda}=\\ =
4\sum_{j\neq i} \left\{\lambda_{ij}\log\frac{\lambda_{ij}}{\Lambda}
+ \frac{\hbar^2}{24N} \left(\frac{K''(\lambda_i)}{K'(\lambda_i)}
- \frac{K''(\lambda_j)}{K'(\lambda_j)}\right) + O(\hbar^4,\Lambda^2)\right\}
\label{QFpexpan}
\ee
In the last line and in the remaining part of the calculation
we neglect all the dependencies on $\gamma=\Lambda^N$,
in this approximation $a_i$ are just the roots $\lambda_i$
of the polynomial
$K(p) = u_N\prod_{i=1}^N (p-\lambda_i)$ and
\be
K'(\lambda_i) = u_N\prod_{j\neq i}\lambda_{ij}, \nn \\
K''(\lambda_i) = 2u_N \sum_{j\neq i}\left( \prod_{k\neq i,j} \lambda_{ik}\right)
\ee
and
\be
\frac{K''(\lambda_i)}{K'(\lambda_i)} =
2\sum_{k\neq i}\frac{1}{\lambda_{ik}}
\ee
Using these formulas, one can check that
(\ref{QFpexpan}) coincides with (\ref{NFpexpan}), provided
$\hbar = \epsilon_1$:
\be
\boxed{
\sum_{j\neq i}\left(\frac{K''(\lambda_i)}{K'(\lambda_i)}
- \frac{K''(\lambda_j)}{K'(\lambda_j)}\right) =
2N \sum_{j\neq i}\frac{1}{\lambda_{ij}}
}
\ee
Indeed,
\be
N=2: & \frac{2}{\lambda_{12}}-\frac{2}{\lambda_{21}} = \frac{4}{\lambda_{12}}, \nn \\
N=3: & 2\cdot\frac{2(\lambda_{12}+\lambda_{13})}{\lambda_{12}\lambda_{13}} -
\frac{2(\lambda_{21}+\lambda_{23})}{\lambda_{21}\lambda_{23}} -
\frac{2(\lambda_{31}+\lambda_{32})}{\lambda_{31}\lambda_{32}} =
6\left(\frac{1}{\lambda_{12}}+\frac{1}{\lambda_{13}}\right), \nn \\
& \ldots
\ee

\section{Conclusion}

In this letter we reported the first check of the claim that 
the (degenerated) Nekrasov functions
are nicely described by the deformation of the SW construction
from quasiclassical to quantum integrable systems
in the simplest non-Abelian case of the $SU(N)$ gauge theory
or the $SL(N)$ affine Toda system.
Switching from the quasiclassical Bohr-Sommerfeld periods
to the exact quantum monodromies preserves consistency of
the SW system of equations, thus, they can be used to define
the deformed prepotential which coincides with Nekrasov's
${\cal F}(\vec a|\epsilon_1)$ with $\epsilon_2=0$.
This seems to be in accordance with the original guess in
\cite{NeSha}.
We performed the check only in the first order, both in
instanton corrections (in $\gamma^2=\Lambda^{2N}$) and
in the quantum deformation parameter $\hbar^2=\epsilon_1^2$,
still this case is already non-trivial.
Of course, higher order corrections deserve to be
found as well.
Generalizations to other models with other gauge groups
and additional matter multiplets,
especially to quiver theories should also be examined.
Of interest is also the similar study of the second deformation
to $\epsilon_1,\epsilon_2\neq 0$ and its relation to another
important hypothesis: the AGT conjecture \cite{AGT}.

\section*{Acknowledgements}

The work was partly supported by Russian Federal Nuclear Energy
Agency and by RFBR grants 07-02-00878 (A.Mir.),
and 07-02-00645 (A.Mor.).
The work was also partly supported
by joint grants 09-02-90493-Ukr,
09-02-93105-CNRSL, 09-01-92440-CE, 09-02-91005-ANF and by Russian President's Grant
of Support for the Scientific Schools NSh-3035.2008.2.


\begin{thebibliography}{12}


\bibitem{NSBS2} A.Mironov and A.Morozov, arXiv:0910.5670

\bibitem{NeSha} N.Nekrasov and S.Shatashvili, arXiv:0908.4052

\bibitem{UFN2} A.Morozov,
Phys.Usp.(UFN) {\bf 37} (1994) 1, hep-th/9303139; hep-th/9502091; hep-th/0502010 \\
A.Mironov, Int.J.Mod.Phys. {\bf A9} (1994) 4355, hep-th/9312212; Phys.Part.Nucl.
{\bf 33} (2002) 537; hep-th/9409190; Theor.Math.Phys. {\bf 114} (1998) 127, q-alg/9711006

\bibitem{SW} N.Seiberg and E.Witten,
Nucl.Phys., {\bf B426} (1994) 19-52, hep-th/9408099;
Nucl.Phys., {\bf B431} (1994) 484-550, hep-th/9407087\\
P.Argyres and A.Shapere, Nucl.Phys., {\bf B461} (1996) 437-459,
hep-th/9509175\\
J.Sonnenschein, S.Theisen and S.Yankielowicz, Phys.Lett.,
{\bf B367} (1996) 145-150, hep-th/9510129

\bibitem{GKMMM} A.Gorsky, I.Krichever, A.Marshakov, A.Mironov, A.Morozov,
Phys.Lett., {\bf B355} (1995) 466-477, hep-th/9505035

\bibitem{Toda} E.Martinec and N.Warner,
Nucl.Phys., {\bf 459} (1996) 97, hep-th/9509161

\bibitem{Cal} 
R.Donagi and E.Witten,
Nucl.Phys., {\bf B460} (1996) 299-334, hep-th/9510101\\
E.Martinec,
Phys.Lett., {\bf B367} (1996) 91-96, hep-th/9510204\\
A.Gorsky, A.Marshakov,
Phys.Lett., {\bf B374} (1996) 218-224, hep-th/9510224\\
H.Itoyama and A.Morozov,
Nucl.Phys., {\bf B477} (1996) 855-877, hep-th/9511126;
Nucl.Phys., {\bf B491} (1997) 529-573, hep-th/9512161; hep-th/9601168

\bibitem{Ruij}
N.Nekrasov,
Nucl.Phys., {\bf B531} (1998) 323-344, hep-th/9609219
\\
H.W. Braden, A.Marshakov, A.Mironov and A.Morozov, Phys.Lett.,
{\bf B448} (1999) 195, hep-th/9812078; Nucl.Phys.,
{\bf B558} (1999) 371, hep-th/9902205

\bibitem{spin}
A.Gorsky, A.Marshakov, A.Mironov, A.Morozov,
Phys.Lett., {\bf B380} (1996) 75-80, hep-th/9603140; hep-th/9604078\\
A.Gorsky, S.Gukov and A.Mironov,
Nucl.Phys., {\bf B517} (1998) 409-461, hep-th/9707120\\
A.Gorsky and A.Mironov, hep-th/0011197

\bibitem{Nek} N.Nekrasov, Adv.Theor.Math.Phys. {\bf 7} (2004) 831-864, hep-th/0206161

\bibitem{LNS}
G.Moore, N.Nekrasov, S.Shatashvili, Nucl.Phys. {\bf B534} (1998) 549-611, hep-th/9711108; 
hep-th/9801061\\
A.Losev, N.Nekrasov and S.Shatashvili, Commun.Math.Phys. {\bf 209} (2000) 97-121, hep-th/9712241; 
ibid. 77-95, hep-th/9803265

\bibitem{DH}
M.Semenov-Tyan-Shansky,
Izv.RAN, ser.Phys. {\bf 40} (1976) 562 \\
J.J.Duisermaat and G.J.Heckman, Inv.Math. {\bf 72} (1983) 153\\
M.Atiyah and R.Bott, Topology, {\bf 23} (1984) 1\\
M.F.Atiyah, Asterisque {\bf 131} (1985) 43\\
E.Witten, Comm.Math.Phys. {\bf 117} (1988) 353;
Int.J.Mod.Phys.{\bf A6} (1991) 2775-2792\\
A.Alekseev, L.Faddeev and S.Shatashvili, J.Geom.Phys. {\bf 1} (1989) 3\\
M.Blau, E.Keski-Vakkuri and A.Niemi, Phys.Lett.{\bf B246}
(1990) 92\\
A.Hietamaki, A.Morozov, A.Niemi and K. Palo,
Phys.Lett. {\bf B263} (1991) 417-424;
Phys.Lett.{\bf B271} (1991) 365-371;
Nucl.Phys.{\bf B377} (1992) 295-338;
Int.J.Mod.Phys. {\bf B6} (1992) 2149-2158


\bibitem{sfst}
A.Gerasimov, S.Khoroshkin, D.Lebedev, A.Mironov and A.Morozov,
Int.J.Mod.Phys. \textbf{A10} (1995) 2589-2614,
hep-th/9405011\\
A.Alexandrov, A.Mironov and A.Morozov,
Int.J.Mod.Phys. {\bf A19} (2004) 4127,
Theor.Math.Phys. {\bf 142} (2005) 349, hep-th/0310113\\
A.Alexandrov, A.Mironov, A.Morozov and P.Putrov, arXiv:0811.2825

\bibitem{MMnf} A.Mironov and A.Morozov, Phys.Lett. {\bf B680} (2009) 188-194,
arXiv:0908.2190

\bibitem{AGT} L.Alday, D.Gaiotto and Y.Tachikawa,
arXiv:0906.3219\\
N.Wyllard, arXiv:0907.2189\\
N.Drukker, D.Morrison and T.Okuda, arXiv:0907.2593\\
A.Marshakov, A.Mironov and A.Morozov, arXiv:0907.3946; Phys.Lett. {\bf B682} (2009) 125-129, 
arXiv:0909.2052; JHEP 11 (2009) 048, arXiv:0909.3338\\
D.Gaiotto, arXiv:0908.0307\\
Andrey Mironov, Sergey Mironov, Alexei Morozov
and Andrey Morozov, arXiv:0908.2064\\
A.Mironov and A.Morozov, Nucl.Phys. {\bf B825} (2009) 1-37 , arXiv:0908.2569; Phys.Lett. {\bf B682} 
(2009) 118-124, arXiv:0909.3531\\
S.Iguri and C.Nunez, arXiv:0908.3460\\
D.Nanopoulos and D.Xie, arXiv:0908.4409; arXiv:0911.1990\\
L.Alday, D.Gaiotto, S.Gukov, Y.Tachikawa and H.Verlinde,
arXiv:0909.0945\\
N.Drukker, J.Gomis, T.Okuda and J.Teschner,
arXiv:0909.1105\\
R.Dijkgraaf and C.Vafa, arXiv:0909.2453\\
R.Poghossian, arXiv:0909.3412\\
A.Gadde, E.Pomoni, L.Rastelli and S.Razamat, arXiv:0910.2225\\
G.Bonelli and A.Tanzini, arXiv:0909.4031\\
L.Alday, F.Benini and Y.Tachikawa, arXiv:0909.4776\\
H.Awata and Y.Yamada, arXiv:0910.4431\\
V.Alba and And.Morozov, arXiv:0911.0363\\
Jian-Feng Wu and Yang Zhou, arXiv:0911.1922

\bibitem{dimamo} S.Kharchev, A.Marshakov, A.Mironov, A.Morozov,
Int. J. Mod. Phys. \textbf{A10} (1995) 2015, hep-th/9312210\\
B.Eynard, J.Stat.Mech. {\bf 0807} (2008) P07023, arXiv:0804.0381\\
A.Klemm and P.Sulkowski, Nucl.Phys. {\bf B819} (2009) 400-430, arXiv:0810.4944 

\bibitem{sygr} N.Nekrasov and A.Okounkov, hep-th/0306238\\
A.Mironov, A.Morozov and S.Natanzon, arXiv:0904.4227

\bibitem{NeShap}
N.Nekrasov, S.Shatashvili, Nucl.Phys., Proc.Suppl. {\bf B192-193} (2009) 91-112, 
arXiv:0901.4744; arXiv:0901.4748

\bibitem{WKB} G.Wentzel, Zeits.f.Physik, {\bf 38} (1926) 518\\
L.Brilloin, Comptes Rendus, {\bf 183} (1926) 24\\
H.A.Kramers, Zeits.f.Physik, {\bf 39} (1926) 828\\
A.Zwaan, Arch.Neerl.des Sciences, {\bf 12} (1929) 33\\
J.L.Dunham, Phys.Rev. {\bf 41} (1932) 713-720

\bibitem{AMM} A.Alexandrov, A.Mironov and A.Morozov, Fortsch.Phys. {\bf 53} (2005) 512-521, 
hep-th/0412205\\
A.Mironov, Theor.Math.Phys. \textbf{146} (2006) 63-72, hep-th/0506158

\bibitem{CIVDV}
R.Dijkgraaf and C.Vafa,
  Nucl.Phys. {\bf B644} (2002) 3, hep-th/0206255;
  Nucl.Phys. {\bf B644} (2002) 21, hep-th/0207106;
hep-th/0208048\\
L.Chekhov and A.Mironov,
  Phys.Lett. {\bf B552} (2003) 293, hep-th/0209085\\
H.Itoyama and A.Morozov,
Prog.Theor.Phys. {\bf 109} (2003) 433-463, hep-th/0212032;
Int.J.Mod.Phys. {\bf A18} (2003) 5889-5906,2003, hep-th/0301136\\
L.Chekhov, A.Marshakov, A.Mironov and D.Vasiliev, 
hep-th/0301071; Proc. Steklov Inst.Math. {\bf 251} (2005) 254,
hep-th/0506075

\bibitem{nfrev} 
R.Flume and R.Pogossian, Int.J.Mod.Phys. {\bf A18} (2003) 2541\\
H.Nakajima and K.Yoshioka, math/0306198, math/0311058\\
S.Shadchin, SIGMA {\bf 2} (2006) 008, hep-th/0601167; hep-th/0502180\\
D.Bellisai, F.Fucito, A.Tanzini and G.Travaglini,
Phys.Lett. {\bf B480} (2000) 365, hep-th/0002110\\
U.Bruzzo, F.Fucito, A.Tanzini, G.Travaglini, Nucl.Phys. {\bf B611} (2001)
205-226, hep-th/0008225\\
U.Bruzzo, F.Fucito, J.Morales and A.Tanzini, JHEP {\bf 0305} (2003) 054,
hep-th/0211108
\\
U.Bruzzo and F.Fucito, Nucl.Phys. {\bf B678} (2004) 638-655, math-ph/0310036\\
F.Fucito, J.Morales and R.Pogossian, JHEP, {\bf 10} (2004) 037, hep-th/040890

\end{thebibliography}
\end{document}